\documentclass[10pt,twoside]{article}
\usepackage{epsf,colordvi,graphicx,color,amsbsy}
\begin{document}

\title{\bfseries
Hysteretic transition from laminar to vortex shedding flow in soap films}
%
\author{ V.K.~Horv\'ath$^*$, J.R.~Cressman, W.I.~Goldburg, and
X.L.~Wu}
\date{}
\maketitle
\vspace*{-9 mm}
{\small
\begin{center}
Department of Physics and Astr., Univ. of
Pittsburgh, Pittsburgh, PA 15260, USA
\end{center}
}
\noindent
$^*$Contact e-mail: vhorvath@pitt.edu
\thispagestyle{empty}

\section{Introduction}
Studies of flow behind a single cylinder have 
greatly contributed to
our understanding of the development of complex flows like
the K\'arm\'an vortex street\cite{karman} or turbulent flows.
At the first instability in such a system
 the laminar flow (LF) becomes time dependent and
vortices appear.
It has been shown that this transition to the vortex shedding phase (VS)
is well described by the supercritical
Hopf bifurcation\cite{provansal_ghia}.

\begin{figure}[!h]
\begin{minipage}[c]{.49 \linewidth}
\includegraphics[width=\linewidth]{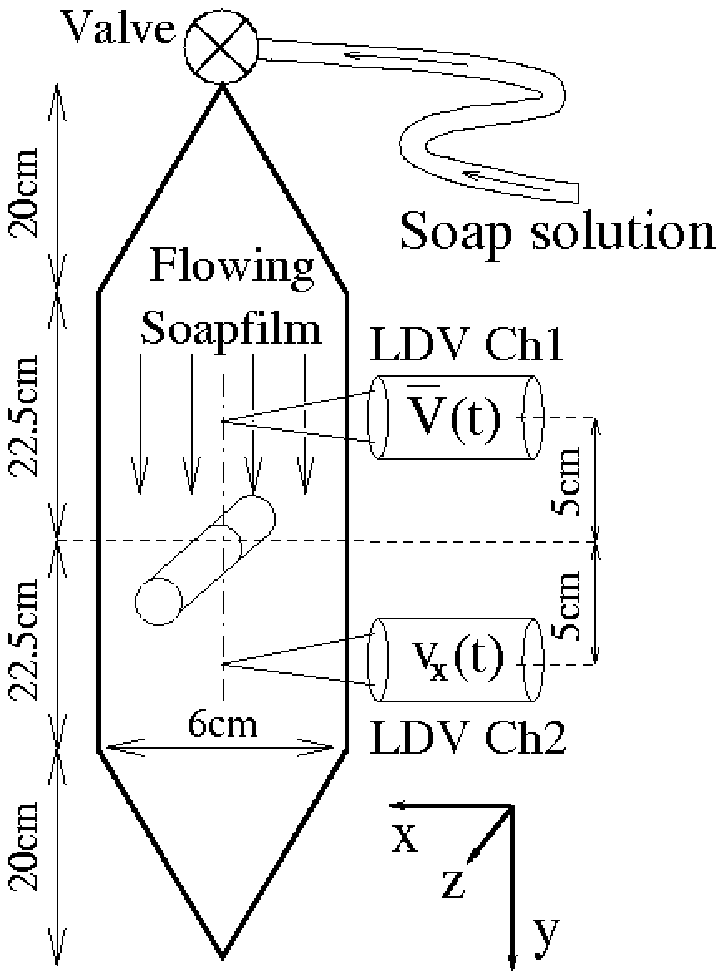}
\caption{Schematic of the experimental setup.
           The diameter $d$ of the rod is 1\/\/mm.
The 0.2\/\/mm thick nylon lines are 
tied to a weight at the bottom.}
\end{minipage}
\hfill
\begin{minipage}[c]{.49 \linewidth}
\centering
\includegraphics[width=\linewidth]{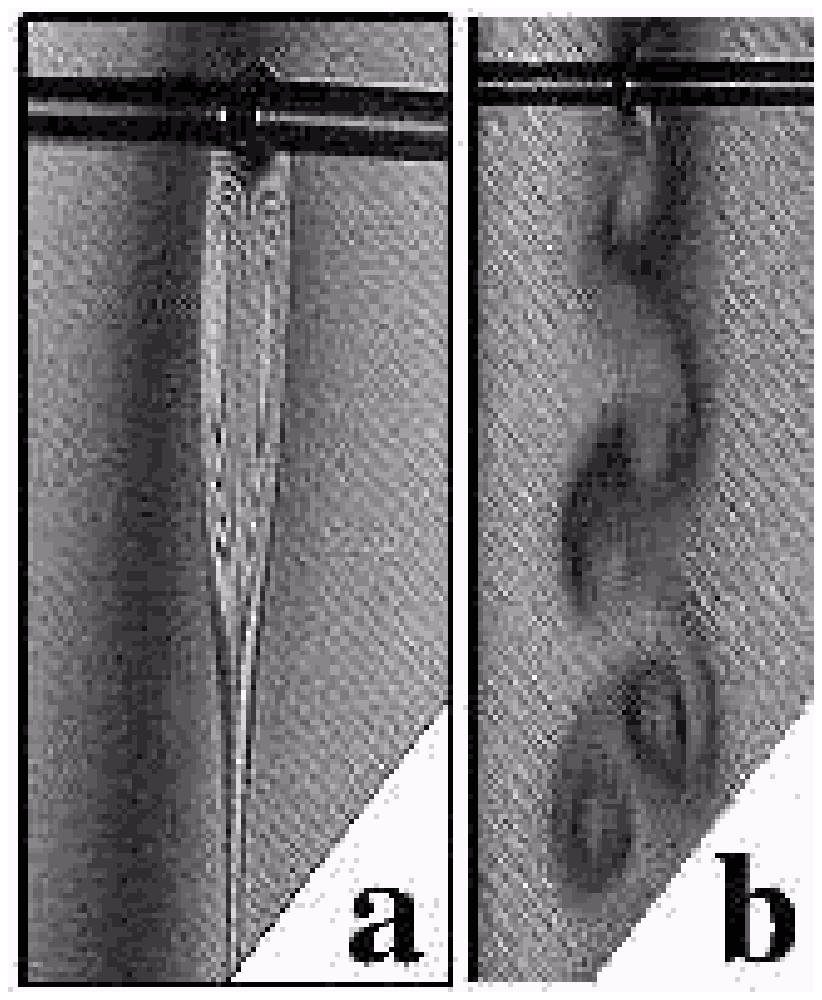}
\caption{Interference pictures of the flowing soap film 
           below the rod in (a) the laminar state and 
           (b) the vortex shedding state.}
\end{minipage}
\vskip-1.0truecm
\end{figure}

\section{Hysteretic Primary Instability in Soap Films}
Soap films have been shown to be a particularly useful 
for the study of two dimensional flows\cite{couder_gharib}.
Here we present measurements demonstrating that the LF$\to$VS transition
can be hysteretic in a (quasi-two dimensional)
soap film penetrated by a glass cylinder\cite{horvath:2000}.
Our experimental setup consists of a
rapidly flowing soap film formed between
two vertically positioned thin nylon lines (see Fig.~1).
The film is fed continuously from the top
through a valve.
To investigate the flow at different Reynolds numbers $Re={\overline
V}d/\nu$
we changed the mean velocity ${\overline V}$ by either opening and closing
the valve at different rates or by changing the separation  of the
nylon lines, using computer controlled stepping motors (here $\nu$ 
is the kinematic viscosity). 
At the center of the parallel
segment of the set-up, a glass rod penetrates the film
in the $z$ direction.
The velocity measurements were made using a dual head
laser Doppler velocimeter (LDV).
The fluctuating  horizontal component $V_{x}(t)$ and $\overline{V}$
were measured simultaneously  below and 
above the rod respectively.

On increasing the flow rate a transition appears from the LF state to 
the VS state.
Below the transition, the
flow is characterized by two counter-rotating vortices
underneath the rod (Fig.~2a). 
In the VS state 
these recirculating vortices peel off the rod and flow
downstream. The continuously generated counter-rotating vortices
form the K\'arm\'an vortex street (Fig.~2b).
All experiments start with slowly increasing 
$\overline{V}$ in the LF regime.
The critical velocity, where VS
commences, is called $V_{c}^{up}$.  
After some waiting time in the VS state, 
$\overline{V}$ is slowly decreased.
At
$V_{c}^{down}$ the system undergoes a reverse transition, i.e.
from VS into LF. 
This cycle was repeated several times in each run with
$V_x(t)$ and $\overline{V}$ being recorded 
simultaneously.

In order to determine the transition velocities, we have
calculated the velocity probability distribution function $P(V_x)$
from $V_{x}(t)$ by a standard binning procedure.  
In Fig.~3 the magnitude of $P(V_x,t)$ is mapped into gray scale values
and is shown as a function of $V_x$ and $t$.
It can be seen
that sharp changes in $P(V_x,t)$ precisely indicate the transition
times $t_d^*$ and $t_u^*$ 
and therefore 
$V_{c}^{down}=\overline{V}(t_d^*)$ and $V_{c}^{up}=\overline{V}(t_u^*)$ too.
It is apparent in
Fig.~3 that $V_c^{up}$ 
is not equal to $V_c^{down}$,
that is, there is no unique critical velocity for the transition.
The Reynolds number 
is roughly 50 in the hysteretic gap.
A large uncertainty in this value comes from the 
poorly determined two-dimensional 
soap film viscosity $\nu$, which depends on the 
film thickness.

\begin{figure}[!h]
\begin{minipage}[c]{.49 \linewidth}
\includegraphics[width=\linewidth]{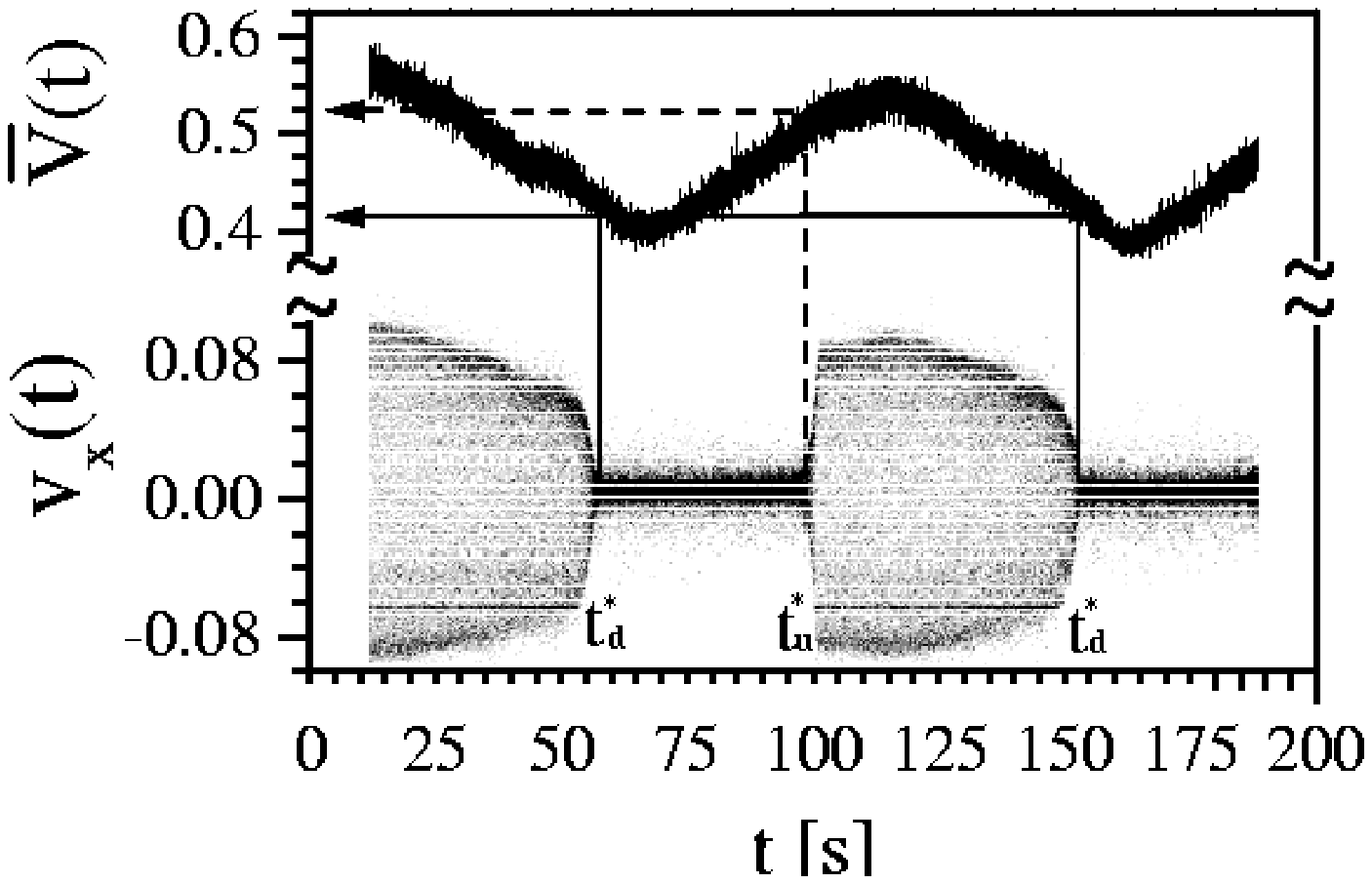}
\caption{Here the function $\overline{V}(t)$ is shown together
     with the $P(v_x,t)$ map at
     $|d\overline{V}/dt|$=0.004 m/s$^2$.
     The gray scale corresponds to the probability density of $v_x$.
     The solid and dashed lines are intended to guide the eye for the
     transitions VS$\to$LF and LF$\to$VS.}
\end{minipage}
\hfill
\begin{minipage}[c]{.49 \linewidth}
\centering
\includegraphics[width=\linewidth]{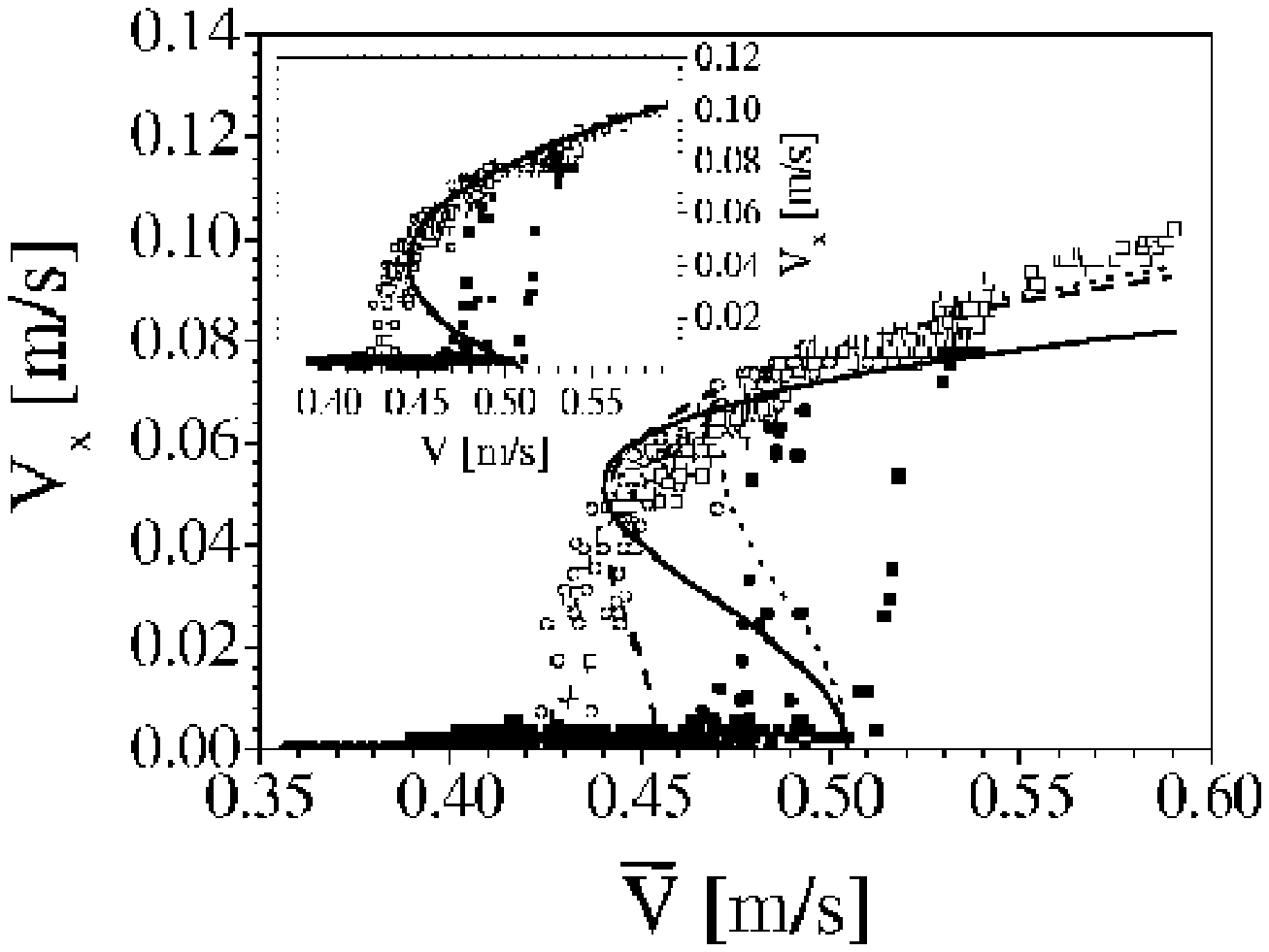}
\caption{This figure shows our  
 experimental data together with different fits
 (see text).
 The circles and squares represent two different experimental run.
 Closed and opened symbols are used for data taken at increasing and
 decreasing $\overline{V}$ respectively. }
\end{minipage}
\end{figure}

It is worth pointing out that 
if we keep the mean flow rate constant in the interval 
$[V_c^{down}$,$V_c^{up}]$, then 
the laminar state can persist for an indefinite length of time.
However applying 
sufficiently large acoustic or pulse-like mechanical perturbations, 
the system could be driven into the VS state. 
The system remains in this state even if the perturbation is turned off.
Applying similar perturbations, VS$\to$LF transition was never observed
at constant $\overline{V}$.
As a result we can conclude that the observed static hysteresis
is {\it not} a result
of some delay in the response of the system to the bifurcation.
Auxiliary experiments were also performed to exclude wetting properties
and the effect 
of the air boundary near the film 
as possible sources of 
hysteresis.

In the absence of any
available theory, we consider 
a fifth order Landau
equation\cite{strogatz} 
$d{\tilde v}(t)/dt={\tilde{a}\tilde v}
   +2{\tilde{b}}v^2{\tilde v}-{\tilde c}v^4{\tilde v} ,
   \label{eqcomplex}$
for the complex velocity
$\tilde v$=$v_{x}+iv_{y}$.
Here all variables denoted by tilde are complex
and $v_i=V_i-<V_i>_t$ ($i$=$x$ or $y$).
The complex velocity 
can also be written as ${\tilde v}$=$ve^{i\phi}$, where
$v$ is the amplitude.
The real part of this equation describes the evolution of $v$.
Its non-trivial positive stationary solution describes the VS state:
\begin{equation}
   {v_{VS}}^2=
  {b_{r}\over c_{r}} ({1+\sqrt{1+a_r\/c_r/b_r^2}},
  \label{v_s}
\end{equation}
where the subscript $r$ is used to designate the  real
part of complex numbers.

Although the application of this model 
for data fitting purposes is not obvious (the relationship between
the parameters $a_r,b_r,c_r$ and the control parameter
of the experiment is unknown), typically 
$a_{r}$ is considered to
be proportional to $(\overline{V}-V_{c}^{up})$.
As one can see in Fig. 4, this generic model of the hysteresis 
is in qualitative agreement with our observations.
However none of our 3 best fits are very satisfactory.   
It is interesting to note that 
a simple three-parameter 
phenomenological equation 
$P_0({\overline V}-V_c)+2P_1v-P_2v^{2}=0$ 
provides a surprisingly good fit to our data in the VS state (see inset
of Fig. 4.). 
Using least square fitting to the solution 
of this equation provides the following result:
$v_x=0.04\pm\sqrt{\overline{V}/39.06-0.0113} $,
where the velocities are in units of $m/s$.

Auxiliary experiments suggest
that the hysteresis may be connected with
the fact that the film may become very thin in the 
recirculating region.
Recognizing that the viscosity of a soap film depends on its 
thickness, the observed hysteresis effect can be qualitatively 
understood; the Reynolds number can now take different values at the 
same value of ${\overline V}$, 
when the parameter is lowered than when it is 
raised.

\section{Conclusions}

\begin{itemize}

\item{We have observed unexpected hysteresis at the onset of the 
      K\'arm\'an vortex street in a quasi two-dimensional soap film.}

\item{The fifth order amplitude equation is in qualitative
agreement with our observations, but a simple phenomenological
equation provides a better numerical fit to the experimental data
in the vortex shedding regime.}

\item{A phenomenological picture is suggested to explain
the origin of the hysteresis.}

\end{itemize}

This material is based upon work supported by the 
NATO under a Grant DGE-9804461 awarded to W.I.~Goldburg and V.K.~Horv\'ath. 
The work was also supported by NSF grant DMR-9622699, 
NASA grant 96-HEDS-01-098 and by additional support
from the Hungarian Science Foundation grant OTKA F17310.

\end{document}